\input lanlmac.tex



\def\unlockat{\catcode`\@=11}
\def\lockat{\catcode`\@=12}

\unlockat

\def\newsec#1{\global\advance\secno by1\message{(\the\secno. #1)}
\global\subsecno=0\global\subsubsecno=0\eqnres@t\noindent
{\bf\the\secno. #1}
\writetoca{{\secsym} {#1}}\par\nobreak\medskip\nobreak}
\global\newcount\subsecno \global\subsecno=0
\def\subsec#1{\global\advance\subsecno
by1\message{(\secsym\the\subsecno. #1)}
\ifnum\lastpenalty>9000\else\bigbreak\fi\global\subsubsecno=0
\noindent{\it\secsym\the\subsecno. #1}
\writetoca{\string\quad {\secsym\the\subsecno.} {#1}}
\par\nobreak\medskip\nobreak}
\global\newcount\subsubsecno \global\subsubsecno=0
\def\subsubsec#1{\global\advance\subsubsecno by1
\message{(\secsym\the\subsecno.\the\subsubsecno. #1)}
\ifnum\lastpenalty>9000\else\bigbreak\fi
\noindent\quad{\secsym\the\subsecno.\the\subsubsecno.}{#1}
\writetoca{\string\qquad{\secsym\the\subsecno.\the\subsubsecno.}{#1}} 

\par\nobreak\medskip\nobreak}

\def\subsubseclab#1{\DefWarn#1\xdef
#1{\noexpand\hyperref{}{subsubsection}%
{\secsym\the\subsecno.\the\subsubsecno}%
{\secsym\the\subsecno.\the\subsubsecno}}%
\writedef{#1\leftbracket#1}\wrlabeL{#1=#1}}
\lockat

\def\IL{\relax{\rm I\kern-.18em L}}
\def\IH{\relax{\rm I\kern-.18em H}}
\def\IR{\relax{\rm I\kern-.18em R}}
\def\IC{\relax\hbox{$\inbar\kern-.3em{\rm C}$}}
\def\IZ{\relax\ifmmode\mathchoice
{\hbox{\cmss Z\kern-.4em Z}}{\hbox{\cmss Z\kern-.4em Z}}
{\lower.9pt\hbox{\cmsss Z\kern-.4em Z}}
{\lower1.2pt\hbox{\cmsss Z\kern-.4em Z}}\else{\cmss Z\kern-.4em
Z}\fi}


\font\manual=manfnt \def\dbend{\lower3.5pt\hbox{\manual\char127}}

\def\IZ{\relax\ifmmode\mathchoice
{\hbox{\cmss Z\kern-.4em Z}}{\hbox{\cmss Z\kern-.4em Z}}
{\lower.9pt\hbox{\cmsss Z\kern-.4em Z}}
{\lower1.2pt\hbox{\cmsss Z\kern-.4em Z}}\else{\cmss Z\kern-.4em
Z}\fi}


\def\IZ{\relax\ifmmode\mathchoice
{\hbox{\cmss Z\kern-.4em Z}}{\hbox{\cmss Z\kern-.4em Z}}
{\lower.9pt\hbox{\cmsss Z\kern-.4em Z}}
{\lower1.2pt\hbox{\cmsss Z\kern-.4em Z}}\else{\cmss Z\kern-.4em
Z}\fi}
\def\IB{\relax{\rm I\kern-.18em B}}
\def\IC{{\relax\hbox{$\inbar\kern-.3em{\rm C}$}}}
\def\ID{\relax{\rm I\kern-.18em D}}
\def\IE{\relax{\rm I\kern-.18em E}}
\def\IF{\relax{\rm I\kern-.18em F}}
\def\IG{\relax\hbox{$\inbar\kern-.3em{\rm G}$}}
\def\IGa{\relax\hbox{${\rm I}\kern-.18em\Gamma$}}
\def\IH{\relax{\rm I\kern-.18em H}}
\def\II{\relax{\rm I\kern-.18em I}}
\def\IK{\relax{\rm I\kern-.18em K}}
\def\IP{\relax{\rm I\kern-.18em P}}

\def\inbar{\,\vrule height1.5ex width.4pt depth0pt}

\font\cmss=cmss10 \font\cmsss=cmss10 at 7pt
\def\IR{\relax{\rm I\kern-.18em R}}


\def\boxit#1{\vbox{\hrule\hbox{\vrule\kern8pt
\vbox{\hbox{\kern8pt}\hbox{\vbox{#1}}\hbox{\kern8pt}}
\kern8pt\vrule}\hrule}}
\def\mathboxit#1{\vbox{\hrule\hbox{\vrule\kern8pt\vbox{\kern8pt
\hbox{$\displaystyle #1$}\kern8pt}\kern8pt\vrule}\hrule}}


\def\inbar{\,\vrule height1.5ex width.4pt depth0pt}

\font\cmss=cmss10 \font\cmsss=cmss10 at 7pt
\def\IR{\relax{\rm I\kern-.18em R}}

\Title{ \vbox{\baselineskip12pt\hbox{hep-ph/9807512}
\hbox{YCTP-P18-98} }}
{\vbox{
\centerline{Modelling Sonoluminescence}
}}\footnote{}
\medskip
\centerline{Alan Chodos $^1$ and Sarah Groff $^2$}
\medskip
\centerline{$^1$Center for Theoretical Physics, Yale University}
\centerline{217 Prospect Street, New Haven, CT 06511-8167 USA}
\medskip
\centerline{$^2$Department of Physics, Yale University}
\centerline{217 Prospect Street, New Haven, CT 06511-8167 USA;}
\centerline{address after September 1, 1998:}
\centerline{Department of Mathematics, MIT, Cambridge, MA 02139 USA}

\bigskip
\bigskip
\centerline{\bf Abstract}
In single-bubble sonoluminescence, a bubble trapped by a sound wave  
in a flask of liquid is forced to expand and contract; exactly once  
per cycle, the bubble emits a very sharp ($< 50 ps$) pulse of  
visible light. This is a robust phenomenon observable to the naked  
eye, yet the mechanism whereby the light is produced is not well  
understood. One model that has been proposed is that the light is  
"vacuum radiation" generated by the coupling of the electromagnetic  
fields to the surface of the bubble. In this paper, we simulate  
vacuum radiation by solving Maxwell's equations with an additional  
term that couples the field to the bubble's motion. We show that, in  
the static case originally considered by Casimir, we reproduce  
Casimir's result. In a simple purely time-dependent example, we find  
that an instability occurs and the pulse of radiation grows  
exponentially. In the more realistic case of spherically-symmetric  
bubble motion, we again find exponential growth in the context of a  
small-radius approximation.

\vfill
\eject

\noindent{\bf I. Introduction}

Single-bubble sonoluminescence [1] is a mysterious phenomenon. A  
small bubble of gas, usually air, is trapped at the center of a  
flask of liquid, usually water,  by the application of an intense  
acoustic field. The frequency of the field is typically 25 or 30  
kHz, and once per cycle, driven by the sound field, the bubble  
undergoes expansion and then rapid contraction. If the parameters  
are right [2] (here "parameters" refers to such things as the  
intensity of the sound field, the concentration of the gas, the  
chemical composition of the gas, the temperature of the water) the  
bubble will emit a very narrow pulse of light during the contraction  
phase of the cycle. The sound wave, with a time scale of tens of  
microseconds, produces a contraction of the bubble measured in tens  
of nanoseconds, which in turn somehow generates a pulse of visible  
light whose duration has recently been measured to be tens of  
picoseconds [3]. Furthermore, even though the motion of the bubble  
is quite violent, if the parameters are right it can be remarkably  
stable, repeating itself over millions of cycles, with the flash of  
light appearing at the same point in the cycle each time [4].

Other aspects of the phenomenology of sonoluminescence are also  
worthy of note. For example, the spectrum of emitted light is only  
partially known, because the water absorbs all wavelengths shorter  
than about 180 nm [5]. The part that is observed looks like the tail  
of a rising distribution, and attempts to fit it to a thermal  
spectrum have led to the speculation that the emitting region is  
very hot, certainly in excess of 25,000 degrees Kelvin, and perhaps  
even as high as a million degrees, at which point nuclear fusion  
might be expected to be significant [6].

Another peculiarity is the fact that, whereas air-filled bubbles  
work well as a vehicle for sonoluminescence, bubbles filled with  
either oxygen or nitrogen, or indeed with a suitable mixture of  
these two gases, do not [7]. The small noble-gas component of air is  
essential for significant sonoluminescence to take place. This  
agrees with the suggestion that, during the first second or so of  
the bubble's oscillation, the oxygen and nitrogen are ionized and  
absorbed by the water, leaving a rarefied bubble filled with noble  
gas [8]. Experiments done with bubbles filled with various noble  
gases confirm that they produce sonoluminescence efficiently.

There is much more to the phenomenology of sonoluminescence, as  
described in a number of recent reviews [9].

Sonoluminescence is a complex phenomenon, involving as it does the  
motion of the bubble, the dynamics of the gas inside the bubble, and  
the mechanism that produces the flash of light. Our main concern in  
this work will be the last of these. The literature contains two  
classes of models to explain the flash of light. One involves the  
gas inside the bubble in an essential way [10]. There is no doubt  
that the gas undergoes compression and heating during the  
contraction phase of the bubble's motion, and this type of  
explanation relies on either thermal radiation, or else  
bremmstrahlung, to produce the light.

The second type of explanation, on which we shall focus, is that the  
observed light is due to "vacuum radiation" [11-15], which is a  
dynamical counterpart to the well-known Casimir effect. In this  
view, given a particular motion of the bubble,

$$
r = R(t) \eqno(1)
$$

\noindent
(here $r$ is the radial coordinate that describes the bubble's  
surface, and $R(t)$ is a prescribed function of time; we assume a  
spherical bubble centered at the origin for simplicity), the  
radiation would take place even if the bubble were completely  
evacuated. The role of the gas, and in particular the special role  
that seems to be played by the noble gases, is merely to modulate  
the motion of the bubble, i.e. to give rise to a specific $R(t)$. It  
is then the motion of the boundary that directly gives rise to the  
radiation.

To explore this idea, our approach will be to take as given all of  
the physics associated with the motion of the bubble and the  
dynamics of the gas, and to extract therefrom the single function  
$R(t)$ which represents the experimentally measured bubble motion.  
Our next task is to construct a model in which the electromagnetic  
field is coupled to the bubble surface at $r = R(t)$.

One approach would be to attempt to derive this coupling from an  
examination of the dielectric properties of water; in practice, this  
would mean simply endowing the water with a dielectric constant  
$\epsilon$, and letting

$$
\epsilon(\vec{x}, t) = \epsilon \theta (r - R(t)) + \theta (R(t) -  
r)  .
$$

Considerable attention in the literature has been devoted to the  
case of $R(t) = const.$, which, although it obviously neglects the  
dynamical mechanism that turns Casimir energy into real photons, is  
supposed to provide an order of magnitude estimate of  the energy  
available, which can then be compared to the energy that is produced  
in sonoluminescence. The question of whether the Casimir energy is  
sufficient in this respect has become a rather controversial one  
[18]. Attempts to treat this problem dynamically have led to  
interesting results, but have not fully resolved the issue  
[12,13,14,19].

In this work, we shall choose a coupling that is not derivable (at  
least by us) from a direct consideration of the underlying physics.   
Rather, the interaction is chosen both for its simplicity and  
because it naturally leads to a coupling localized on the boundary  
$r = R(t)$. In addition, as well shall show below, when one  
considers the case of two static parallel plates (Casimir's original  
problem) one recovers precisely the original Casimir energy, and is  
therefore encouraged to hope that the model may be a valid  
representation of the dynamical situation as well.

In section 2 we shall introduce the model and derive the boundary  
conditions on $\vec{E}$ and $\vec{B}$ that it implies. In section 3,  
we look at two instructive cases that are not directly related to  
sonoluminescence: the case of static, parallel plates mentioned  
above, and the case of a strictly time-dependent source, with no  
spatial dependence. In this latter case, we shall discover the  
existence of unstable modes that can lead to production of radiation  
at unexpectedly large rates.

In section 4, we tackle the case of greatest interest, the  
collapsing bubble. Even classically, we are unable to solve the  
equations exactly, but we develop an approximation scheme that  
relies on the fact that the radius of the bubble is small, in the  
sense that $R(t) << c T$, where $T$ is a time characteristic of the  
width of the sonoluminescent pulse. In this approximation we find  
the same sort of unstable modes that existed in the purely  
time-dependent case. Section 5 is devoted to conclusions, and we  
have also included an appendix in which further properties of the  
$F\tilde{F}$ interaction term are discussed.

\bigskip\noindent
{\bf II.  The Model}

We consider the following Lagrange density [16]:

$$
{\cal L} = - {1 \over 4} [F_{\mu\nu} F^{\mu\nu} + f(x) F_{\mu\nu}  
\tilde{F}^{\mu\nu}]~~ . \eqno(2)
$$

\bigskip\noindent
Here $F_{\mu\nu}$ has its usual meaning: ~$F_{\mu\nu} =  
\partial_{\mu} A_{\nu} - \partial_{\nu} A_{\mu}$ where $A_{\mu}$ is  
the 4-vector electromagnetic potential, and $\tilde{F}_{\mu\nu} = {1  
\over 2} \epsilon_{\mu\nu\rho\sigma} F^{\rho\sigma}$, where  
$\epsilon_{\mu\nu\rho\sigma}$ is the totally antisymmetric symbol on  
4 indices, and $\epsilon^{0123} = 1$.

As is well-known, $F_{\mu\nu} \tilde{F}^{\mu\nu}$ is a total  
divergence:

$$
F_{\mu\nu} \tilde{F}^{\mu\nu} = \partial_{\mu}  
[2\epsilon^{\mu\nu\rho\sigma} A_{\nu} \partial_{\rho} A_{\sigma}]  
\eqno(3)
$$

\bigskip\noindent
so up to a surface term, ${\cal L}$ can be written

$$
{\cal L} = - {1 \over 4} [F_{\mu\nu} F^{\mu\nu} - 2(\partial_{\mu}f)  
e^{\mu\nu\rho\sigma} A_{\nu} \partial_{\rho} A_{\sigma}] ~~.  
\eqno(4)
$$

\bigskip\noindent
Hence, by choosing

$$f(x) = f_0 \theta [g(x)] \eqno(5)
$$

\bigskip\noindent
we obtain $\partial_{\mu}f = f_0 \partial_{\mu}g \delta(g)$ so the  
second term in ${\cal L}$ represents the coupling of the  
electromagnetic field to the surface given by $g(x) = 0$. ($f_0$ is  
a dimensionless constant.) For the case of the sonoluminescing  
bubble, we would choose $g(\vec{x}, t) = r - R(t)$.

The equations of motion that follow from ${\cal L}$ are

$$
\partial_{\mu}[F^{\mu\nu} + f(x) \tilde{F}^{\mu\nu}] = 0 ~~,  
\eqno(6)
$$

\bigskip\noindent
or, since $\partial_{\mu} \tilde{F}^{\mu\nu} = 0$ identically,

$$
\partial_{\mu} F^{\mu\nu} + (\partial_{\mu}f) \tilde{F}^{\mu\nu} = 0  
~~. \eqno(7)
$$

\bigskip\noindent
If we define $\vec{E}$ and $\vec{B}$ in the usual way, we obtain the  
modified Maxwell equations

$$
\vec{\bigtriangledown} \cdot \vec{E} + \vec{\bigtriangledown}f \cdot  
\vec{B} = 0 \eqno(8)
$$

\medskip
$$
\vec{\bigtriangledown} \times \vec{B} - \vec{\dot{E}} -  
\dot{f}\vec{B} - \vec{\bigtriangledown}f \times \vec{E} = 0 \eqno(9)
$$

\bigskip\noindent
together with $\vec{\bigtriangledown} \cdot \vec{B} = 0$ and  
$\vec{\bigtriangledown} \times \vec{E} + \vec{\dot{B}} = 0$. With  
the choice (5), we have $\dot{f} = f_0 n^0 \delta (g)$ and  
$\vec{\bigtriangledown}f = f_0 \vec{n} \delta (g)$ where $n_{\mu} =  
(\dot{g}, \vec{\bigtriangledown}g)$ is the four-normal to the  
surface. To see what these equations entail, we write

$$
\vec{E} = \vec{E}_1 \theta(g) + \vec{E}_2 \theta(-g) \eqno(10a)
$$
$$\vec{B} = \vec{B}_1 \theta(g) + \vec{B}_2 \theta(-g) \eqno(10b)
$$

\bigskip\noindent
and substituting into equations (8) - (9) , we find that the pair  
$(\vec{E}_1, \vec{B}_1)$ satisfy the free Maxwell equations for $g >  
0$, and likewise $(\vec{E}_2, \vec{B}_2)$ satisfy them for $g < 0$.  
At $g = 0$, we have the boundary conditions:

$$
\vec{n} \cdot (\vec{E}_1 - \vec{E}_2) + f_0 \vec{n} \cdot \vec{B} =  
0 \eqno(11)
$$

$$
\vec{n} \times (\vec{B}_1 - \vec{B}_2) - n_0 (\vec{E}_1 - \vec{E}_2)  
- f_0(n_0 \vec{B} + \vec{n} \times \vec{E}) = 0 \eqno(12)
$$

$$
\vec{n} \cdot (\vec{B}_1 - \vec{B}_2) = 0 \eqno(13)
$$

$$
\vec{n} \times (\vec{E}_1 - \vec{E}_2) + n_0 (\vec{B}_1 - \vec{B}_2)  
= 0 ~~. \eqno(14)
$$

\bigskip\noindent
Notice that the second pair of equations, (13) - (14), removes the  
ambiguity as to which values of $\vec{B}$ and $\vec{E}$ to use in  
the terms proportional to $f_0$ in the first pair of equations, (11)  
- (12).

\bigskip\noindent
{\bf III. ~Special Cases}

\medskip\noindent
{\it (a) Parallel plates}

Before dealing with the time-dependent case, let us explore the  
physical significance of our model by revisiting the case originally  
considered by Casimir [17], i.e. two infinite parallel planes  
separated by a distance $a$. We take $g(\vec{x}) = z(z - a)$. The  
planes divide space into 3 regions, $z < 0$, $0 < z < a$ and $z >  
a$, and in each region we can choose a plane-wave solution to  
Maxwell's equations:

$$
\vec{E} = e^{-i\omega t} [\vec{e} e^{i\vec{k} \cdot \vec{x}} +  
\vec{e}^{~\prime} e^{i\vec{k}^{\prime} \cdot \vec{x}}] \eqno(15)
$$

\bigskip\noindent
where $\vec{k} = (k_1, k_2, k_3)$ and $\vec{k}^{\prime} = (k_1, k_2,  
- k_3)$ and

$$\vec{B} = e^{-i\omega t} [\vec{b} e^{i\vec{k} \cdot \vec{x}} +  
\vec{b}^{\prime} e^{i\vec{k}^{\prime} \cdot \vec{x}}] ~.\eqno(16)
$$

\bigskip\noindent
We need both $\vec{k}$ and $\vec{k}^{\prime}$ because the boundary  
conditions will mix them.

Maxwell's equations imply that

$$
\vec{k} \times \vec{b} = - \omega  \vec{e}  
\quad\quad\quad\quad\quad\quad \vec{k} \times \vec{e} = \omega   
\vec{b} \eqno(17)
$$
\centerline{and}
$$
\vec{k}^{\prime} \times \vec{b}^{\prime} = - \omega   
\vec{e}^{~\prime} \quad\quad\quad\quad\quad\quad \vec{k}^{\prime}  
\times \vec{e}^{~\prime} = \omega  \vec{b}^{\prime} \eqno(18)
$$

\bigskip\noindent
in each of the three regions, which in turn imply the $\omega^2 =  
\vec{k}^2 = \vec{k}^{\prime 2}$.

It is now a matter of implementing the boundary conditions at $z =  
0$ and at $z = a$. After some algebra, it is not hard to show that  
the content of these conditions reduces to

$$
z = 0 : ~~~~b_3 + b^{\prime}_3 = 0 \eqno(19)
$$
$$
z = a : ~~~~b_3 e^{ik_3a} + b^{\prime}_3 e^{-ik_3a} = 0 \eqno(20)
$$

\bigskip\noindent
(here $b_3$ means the third component of $\vec{b}$ in the region $0  
< z < a$, and similarly for $b_3^{\prime}$) from which it follows  
that

$$
k_3a = n\pi , ~~~ n = 0, \pm 1, \pm 2, \cdot\cdot\cdot\cdot ~~.  
\eqno(21)
$$

\bigskip\noindent
This is exactly the same spectrum used by Casimir in his original  
paper, and therefore the Casimir energy $\delta E$ will be the same  
as his result:

$$
{\delta E \over L^2} = - {\pi^2 \over 720a^3} \eqno(22)
$$

\noindent
where $L^2$ is the area of one of the plates.

\medskip\noindent
{\it (b) Time dependent source}

Armed with the knowledge that our model reproduces the static  
Casimir energy, we now proceed to another simple example, which is  
very different physically: ~we take $f(\vec{x}, t)$ to depend only  
on $t$:

$$
\eqalign{
f(x, t) &= 0, ~~~~~t \leq 0 \cr
&= gt, ~~~~0 < t < T \cr
&=gT, ~~~~t \geq T ~.  \cr
}
\eqno(23)$$

\bigskip\noindent
Since $f$ is not a step function, there is no bubble in this case.  
Rather, the source is turned on everywhere at once at $t = 0$, and  
is turned off again at $t = T$ ($f =$ const. is without physical  
consequence, because $F_{\mu\nu} \tilde{F}^{\mu\nu}$ is a total  
divergence).

The equations of motion in this case are of course just  the free  
Maxwell equations for $t < 0$ and $t > T$, whereas for $0 < t < T$,  
one of the Maxwell equations is modified:

$$
\vec{\bigtriangledown} \times \vec{B} - \vec{\dot{E}} = g\vec{B} ~.  
\eqno(24)
$$

\bigskip\noindent
One can now study a plane-wave solution, propagating, say, along the  
$z$-axis. For $t < 0$ we write

$$
\eqalign{
\vec{B} &= (a \hat{x} + b \hat{y}) e^{i(kz - \omega t)} \cr
\vec{E} &= (b \hat{x} - a \hat{y}) e^{i(kz - \omega t)}, \cr
}
\eqno(25)$$

\bigskip\noindent
with $k^2 = \omega^2$. At $t = 0$, this will be matched to a  
solution of the form

$$
\eqalign{
\vec{B} &= e^{ikz} \biggr{[} e^{-i\Omega t} (\alpha \hat{x} + \beta  
\hat{y}) + e^{i\Omega t} (\gamma \hat{x} + \delta \hat{y}) \biggr{]}  
\cr
\vec{E} &= e^{ikz} \biggr{[} e^{-i\Omega t} (\beta \hat{x} - \alpha  
\hat{y}) + e^{i\Omega t} (\delta \hat{x} - \gamma \hat{y}) \biggr{]}  
\cr
}
\eqno(26)$$

\bigskip\noindent
where, because of the modification to Maxwell's equations, one has  
$\Omega^2 = \Omega^2_{\pm} = [k (k \pm g)]$.

Corresponding to each of these solutions is a particular  
polarization $\hat{C}_{\pm} = {1 \over \sqrt{2}} [\hat{x} \mp i  
\hat{y}]$. When matched to the $t < 0$ solution, the expression for  
$\vec{B}$, $0 < t < T$, becomes

$$
\eqalign{
e^{-ikz} \vec{B} &= ({k + \Omega_+ \over 2 \Omega_+}) ~{a + ib \over  
\sqrt{2}} \hat{C}_+ e^{-i\Omega_+t} - ({k - \Omega_+ \over  
2\Omega_+}) ~{a + ib \over \sqrt{2}} \hat{C}_+ e^{i\Omega_+t} \cr
&+ ({k + \Omega_- \over 2 \Omega_-}) ~{a - ib \over \sqrt{2}}  
\hat{C}_- e^{-i\Omega_-t} - ({k - \Omega_- \over 2\Omega_-}) ~{a -  
ib \over \sqrt{2}} \hat{C}_- e^{i\Omega_-t} ~,\cr
}
\eqno(27)$$

\bigskip\noindent
with a similar expression for $\vec{E}$.

One can extend this analysis by matching this solution to a suitable  
expression for $\vec{E}$ and $\vec{B}$ in the region $t > T$, where  
of course $\omega^2 = k^2$ again. But we shall not need this  
extension in what follows.

The feature most worthy of note is that, (for $g > 0$) the frequency  
$\Omega_-$ becomes imaginary when $k < g$. (If $g < 0$, then  
$\Omega_+$ becomes imaginary.) Hence $\vec{E}$ and $\vec{B}$ grow  
exponentially with time over the interval $0 < t < T$. We shall see  
below that, at least in a certain approximation, this feature  
persists in the case of a spherically oscillating bubble.

To quantize this model, we can express $\vec{E}$ and $\vec{B}$ in  
terms of a vector potential $\vec{A}$, and endow the fourier  
coefficients of $\vec{A}$ with the appropriate commutation  
relations. Effectively this means that the coefficients $a$ and $b$  
in the above expressions become quantum operators. We must also  
generalize our solution to the case of a plane wave propagating in  
an arbitrary direction, but this is easily done since the $z$-axis  
used above was in no way special.  It is of interest to compute the  
rate of energy production per unit volume by the external source. We  
do this by forming the hamiltonian density

$$
{\cal H} = {1 \over 2} (\vec{E} \cdot \vec{E} + \vec{B} \cdot  
\vec{B}) \eqno(28)
$$

\bigskip\noindent
and taking the vacuum expectation value of its time derivative.  
${\cal H}$ is normal-ordered so that $\langle 0 \mid {\cal H} \mid 0  
\rangle = 0$ for $t < 0$. Since the expressions are quite lengthy,  
we simplify matters by retaining only those pieces which grow  
exponentially.  After some calculation, we then find

$$
{d \over dt} \langle 0 \mid {\cal H} \mid 0 \rangle_{exp.} =  
\theta(t) \theta(T - t) {g^5 \over 16 \pi^2} \int_0^1 ~{x^{5/2}  
\over (1 - x)^{1/2}} ~e^{2gt\sqrt{x(1 - x)}} dx \eqno(29)
$$

\bigskip\noindent
where the notation "exp" on the matrix element means the  
exponentially growing piece. A simple stationary-phase estimate of  
the integral gives

$$
{d \over dt} \langle 0 \mid {\cal H} \mid 0 \rangle_{exp} \simeq  
\theta(t) \theta(T - t) {g^5 \over 64 \pi^2} ~e^{gt} ~. \eqno(30)
$$

\bigskip\noindent
We can try to connect this to sonoluminescence (despite the fact  
that there is no bubble) by choosing $T = 10^{-11}$ sec (the  
duration of a typical pulse) and $1/g = 2 \times 10^{-7}$ m. (the  
cutoff on the observed spectrum). We then find $gT = 1.6 \times  
10^4$, which, needless to say, produces a huge number  when inserted  
in the exponent in eq. (30).

At this point, we can simply argue that our model is too far removed  
from the phenomenology of sonoluminscence to be expected to give  
reasonable results. Later, however, we shall have to deal with this  
question in the context of a more realistic model, to which we now  
turn.

\vfill\eject
\noindent
{\bf IV. ~The collapsing bubble}

We take $f(x) = f_0 \theta(r - R(t))$. Our strategy will be to  
attempt to solve the classical problem, looking for the kind of  
exponential behavior in time that we found in the previous example.   
If this is indeed found, then, reasoning by analogy with the  
previous example, we will argue that, when quantized, the model will  
produce an exponentially growing pulse of vacuum radiation over some  
period of time.

Because of the spherical symmetry, it is appropriate to expand  
$\vec{E}$ and $\vec{B}$ in terms of vector spherical harmonics.  
Different values of $l$ and $m$ will not couple to each other.

We write

$$
\vec{E} = e_1 \vec{L} Y_{lm} + (r e_2^{\prime} + e_2)  
\vec{\bigtriangledown} Y_{lm} + {l(l + 1) \over r^2} e_2 \vec{r}  
Y_{lm} \eqno(31)
$$

\noindent
and

$$
\vec{B} = b_1 \vec{L} Y_{lm} + (r b_2^{\prime} + b_2)  
\vec{\bigtriangledown} Y_{lm} + {l(l + 1) \over r^2} b_2 \vec{r}  
Y_{lm} ~. \eqno(32)
$$

\bigskip\noindent
Here $\vec{L} = {1 \over i} \vec{r} \times \vec{\bigtriangledown}$,  
and the $e$'s and $b$'s are functions of $r$ and $t$; we cannot  
separate variables  any further because the boundary conditions will  
mix $r$ and $t$.

These forms automatically satisfy $\vec{\bigtriangledown} \cdot  
\vec{E} = \vec{\bigtriangledown} \cdot \vec{B} = 0$. The rest of  
Maxwell's equations imply that

$$
\ddot{\varphi} - \varphi^{\prime\prime} - {2 \over r}  
\varphi^{\prime} + {l(l + 1) \over r^2} \varphi = 0 ~. \eqno(33)
$$

\bigskip\noindent
where $\varphi$ is any of the $e$'s or $b$'s, and furthermore that

$$
e_1 = i \dot{b}_2 ~~{\rm and} ~~ b_1 = -i \dot{e}_2 ~~. \eqno(34)
$$

\bigskip\noindent
In these equations, a prime is ${\partial \over \partial r}$ and an  
overdot is ${\partial \over \partial t}$. So if we cast the boundary  
conditions entirely in terms of $b_2$ and $e_2$, we can recover  
$e_1$ and $b_1$ from equation (34).

In fact it is not hard to express the boundary conditions in terms  
of $e_2$ and $b_2$.  We expand $\vec{E}$ and $\vec{B}$ as in eqs.  
(31) and (32) separately inside and outside the bubble, and we let  
$\Delta\varphi = \varphi_{out} - \varphi_{in}$ where once again  
$\varphi$ is any of the $e$'s or $b$'s. Then we find, at $r = R(t)$:

$$\Delta b_2 = 0 ~; \eqno(35)
$$

$$
\Delta e_2 = - f_0b_2 \eqno(36)
$$

$$
\Delta\dot{b}_2 = {- f_0 \dot{R} \over 1 - \dot{R}^2} ~[{e_2 \over  
R} + e^{\prime}_2 + \dot{R} \dot{e}_2] \eqno(37)
$$

\noindent
and

$$
\Delta\dot{e}_2 = {- f_0 \over 1 - \dot{R}^2} ~[{\dot{R} b_2 \over  
R} + \dot{b}_2 +  \dot{R} b^{\prime}_2] ~. \eqno(38)
$$

\bigskip\noindent
One can show that the expressions on the right-hand side of these  
equations all have zero discontinuity at $r = R(t)$, so there is no  
ambiguity as to which values to insert.

For simplicity, we choose to analyze the case $l = 1$. Then the most  
general solutions to equation (33), for $b_2$ and $e_2$, are:

$$
\eqalign{
b_2^{out} &= {\partial \over \partial r} ~[{1 \over r}  
~(\tilde{\beta}(t + r) - \hat{\beta}(t - r))] \cr
b_2^{in} &= {\partial \over \partial r} ~[{1 \over r} ~(\beta(t + r)  
- \beta(t - r))] \cr
e_2^{out} &= {\partial \over \partial r} ~[{1 \over r}  
~(\tilde{\gamma}(t + r) - \hat{\gamma} (t - r))] \cr
{\rm and}  ~~~~e_2^{in} &= {\partial \over \partial r} ~[{1 \over r}  
~(\gamma (t + r) - \gamma (t - r))] ~. \cr
}
\eqno(39)
$$

\bigskip\noindent
Here the $\beta$'s and $\gamma$'s are arbitrary functions of the  
indicated arguments.  In writing these equations,  we have imposed  
the requirement that $b_2$ and $e_2$ be regular at $r = 0$. The  
functions $\tilde{\beta}$ and $\tilde{\gamma}$ determine the waves  
propagating inward from infinity and should be taken as initial  
data. In principle, it should be possible to use the boundary  
conditions, eqns. (35) - (38), to determine the inside solution,  
specified by $\beta$ and $\gamma$, and the outgoing waves specified  
by $\hat{\beta}$ and $\hat{\gamma}$. The effect we are looking for  
is to see whether non-exponentially growing incoming data  
$\tilde{\beta}$ and $\tilde{\gamma}$ can generate exponential growth  
in the outgoing solution $\hat{\beta}$ and $\hat{\gamma}$.

Unfortunately, when we substitute the forms (39) into the boundary  
conditions, we find rather complicated functional difference  
equations that we do not know how to solve. Instead we rely on an  
approximation that is based on the following observation.  
Phenomenologically, the smallest time scale we are interested in is  
the width of the sonoluminescent pulse, $\Delta t_{min} \cong  
10^{-11}$ sec. The largest length scale we are interested in is the  
maximum size of the bubble, $R_{max} \cong 10^{-4}$ m. But in units  
with $c = 1$,

$$
{R_{max} \over \Delta t_{min}} = 10^7 m/sec = {1 \over 30} ~.  
\eqno(40)
$$

\bigskip\noindent
(this is actually a generous overestimate, since sonoluminescence  
occurs when the bubble radius is at least an order of magnitude  
smaller than $R_{max}$).  Thus it might make sense to regard $R(t)$  
as a small parameter, and to expand our equations accordingly. We do  
this as follows: ~in the expressions for $b_2$ and $e_2$, equation  
(39), we replace the arguments $t \pm r$ by $t \pm \epsilon r$,  
where $\epsilon$ is a bookkeeping parameter in which we perform a  
systematic expansion. At the end we set $\epsilon = 1$.

To obtain a consistent expansion, we not only expand the arguments  
of the functions, we must also expand the functions themselves:

$$
\varphi(t) = \varphi_0(t) + \epsilon \varphi_1(t) + \epsilon^2  
\varphi_2(t) + \cdots \eqno(41)
$$

\bigskip\noindent
where $\varphi$ stands for any of the unknowns, $\hat{\beta},  
\hat{\gamma}, \beta$ and $\gamma$. The input functions  
$\tilde{\beta}$ and $\tilde{\gamma}$ are regarded as known and are  
not so expanded. The advantage of this expansion procedure is that  
we obtain thereby relations among functions all of which are  
evaluated at the same argument $t$. Henceforth we denote ${d \over  
dt}$ by a prime.

To obtain non-trivial results, we must retain terms up to order  
$\epsilon^3$. For convenience, we introduce the notation $\rho_1(t)  
= \beta_0^{\prime\prime\prime}(t)$ and $\rho_2(t) =  
\gamma_0^{\prime\prime\prime}(t)$. We find

$$
\eqalign{
\hat{\beta}_0 &= \tilde{\beta} ~; ~~~~\hat{\gamma}_0 =  
\tilde{\gamma} \cr
\hat{\beta}_1 &= \hat{\beta}_2 = \hat{\gamma}_1 = \hat{\gamma}_2 = 0  
~.\cr
\hat{\beta}_3 &= {2 \over 3} R^3 ~[\rho_1 -  
\tilde{\beta}^{\prime\prime\prime}] \cr
\hat{\gamma}_3 &= {2 \over 3} R^3 ~[\rho_2 -  
\tilde{\gamma}^{\prime\prime\prime} - f_0 \rho_1] ~.\cr
} \eqno(42)
$$

\noindent
and

$$
\eqalign{
\rho_1 &= \tilde{\beta}^{\prime\prime\prime} - {1 \over 3} ~({f_0  
\over 1 - \dot{R}^2}) ~[2 \rho_2 + R\dot{R} \rho_2^{\prime}] \cr
\rho_2 &= \tilde{\gamma}^{\prime\prime\prime} - {1 \over 3} ~({f_0  
\over 1 - \dot{R}^2}) ~[(1 - 3\dot{R}^2) ~\rho_1 + R\dot{R}  
\rho_1^{\prime}] ~. \cr
} \eqno(43)
$$

\bigskip\noindent
The functions whose behavior we want to study are $\hat{\beta}_3$  
and $\hat{\gamma}_3$, which give the first non-trivial corrections  
to the outgoing waves $\hat{\beta}$ and $\hat{\gamma}$. Our strategy  
will be to solve eqn. (43) for $\rho_1$ and $\rho_2$, and then  
evaluate $\hat{\beta}_3$ and $\hat{\gamma}_3$ from eqn. (42). If we  
make the further approximation $\dot{R}^2 << 1$, which must surely  
be true for any realistic bubble motion, we can drop the $\dot{R}^2$  
terms on the r-hs of eqn. (43), which can then be rewritten as

$$
{- ~f_0 R\dot{R} \over 3} ~\biggr{(} \matrix{
\rho_1^{\prime}\cr
\rho_2^{\prime}\cr
}
\biggr{)} = M \biggr{(} \matrix{
\rho_1\cr
\rho_2\cr
}
\biggr{)} + \biggr{(}\matrix{
\tilde{\gamma}^{\prime\prime\prime}\cr
\tilde{\beta}^{\prime\prime\prime}\cr
}
\biggr{)} ~. \eqno(44)
$$

\bigskip\noindent
where $M$ is the matrix $\pmatrix{
{1 \over 3} f_0 & -1 \cr
-1 & {- 2~ \over 3} f_0 \cr
}$. The eigenvalues of $M$ are

$$\lambda_{\pm} = {1 \over 2} ~[\pm \sqrt{f_0^2 + 4} - {1 \over 3}  
f_0] ~. \eqno(45)
$$

\bigskip\noindent
Note that $\lambda_+ > 0, \lambda_- < 0$. Let $\theta$ be the  
orthogonal matrix

$$
\theta = {1 \over \sqrt{2}} \left [\matrix{
k_+ & - k_- \cr
k_- & k_+ \cr
}\right ]
\eqno(46)$$

\bigskip\noindent
where $k_{\pm} = [ 1 \pm {f_0 \over \sqrt{f_0^2 + 4}}]^{1/2}$. Then,  
setting $\left (\matrix{
\sigma_1\cr
\sigma_2\cr
}\right ) = \theta \left (\matrix{
\rho_1\cr
\rho_2\cr
}\right )$, our equation becomes

$$
{- f_0 R\dot{R} \over 3} \left (\matrix{
\sigma_1^{\prime}\cr
\sigma_2^{\prime}\cr
}\right ) = \pmatrix{
\lambda_+ & \sigma_1 \cr
\lambda_- & \sigma_2 \cr
} + \left (\matrix{
\kappa_1\cr
\kappa_2\cr
}\right )
\eqno(47)$$

\bigskip\noindent
where $\left (\matrix{
\kappa_1\cr
\kappa_2\cr
}\right )
= \theta \left (\matrix{
\tilde{\gamma}^{\prime\prime\prime}\cr
\tilde{\beta}^{\prime\prime\prime}\cr
}\right )
$. Thus we have to solve

$$
{- f_0 R\dot{R} \over 3} ~\sigma^{~\prime} = \lambda \sigma + \kappa  
\eqno(48)
$$

\bigskip\noindent
whose solution is

$$
\sigma(t) = exp [- \int_{t_0}^t dt^{~\prime} ~{3 \lambda \over f_0   
R\dot{R}}] ~\biggr{[}- \int_{t_0}^t dt^{~\prime}  
~exp~\{\int_{t_0}^{t^{\prime}} dt^{~\prime\prime} {3 \lambda \over  
f_0 R\dot{R}}\} ({3 \kappa \over f_0 R\dot{R}})  + \sigma  
(t_0)\biggr{]}. \eqno(49)
$$

\bigskip\noindent
There are two such solutions, one for $\lambda_+$ and one for  
$\lambda_-$. Generally, the exponential behavior that is manifest on  
the $r.h.s.$ of eqn. (49) will cancel in the first term, but will  
survive in the term proportional to $\sigma ({t_0})$. Because in the  
region of interest we have $R\dot{R} < 0$, it will be $\lambda_+$  
that gives the exponentially growing behavior (for $f_0 > 0$).

How do we interpret this result? First, we must recognize that the  
approximation we have made is potentially very dangerous, because  
the highest derivative in eqn. (48) is multiplied by a factor, ${-  
f_0 R\dot{R} \over 3}$, which we expect to be quite small, and  
indeed which we expect to go to zero for $\mid t \mid$ large. [Here  
we are restricting ourselves to only one cycle of the bubble's  
motion, so we take $R(t) \rightarrow const.$ as $\mid t \mid  
\rightarrow \infty$.] As a consequence, it appears from the solution  
that the exponential behavior becomes more pronounced the smaller  
$R\dot{R}$ becomes, whereas we know from the original equation,  
(48), that for $R\dot{R}$ strictly zero the solution is just $\sigma  
= - {1 \over \lambda} \kappa$, which exhibits no exponential  
behavior at all.

Our response to this is to imagine that for $\mid f_0 R\dot{R} \mid$  
below some threshold value, it is indeed negligible, and therefore  
$\sigma = - {1 \over \lambda} \kappa$. At some time $t_0$, $\mid f_0  
R\dot{R} \mid$ crosses the threshold, and the solution (49) kicks  
in. We therefore fix the arbitrary constant $\sigma(t_0)$ to be $-  
{1 \over \lambda} \kappa (t_0)$. As we have already observed, for  
one of the two choices of $\lambda$, $\sigma$ will be an  
exponentially growing function (except for the unlikely possibility  
that $\kappa(t_0) = 0$).

The good news is that, within the context of our approximation, we  
have found the exponential behavior that we are looking for. The bad  
news is that, just as in the earlier, simpler example, we have an  
embarrassment of riches. There appears to be no mechanism within the  
model for turning the exponential behavior off - we have already  
fixed the one free parameter $\sigma(t_0)$. Let us assume that there  
is a dissipative mechanism, having to do with the properties of the  
gas inside the bubble or the liquid outside it, that 

we should add to our model. This will abort the vacuum radiation  
after some characteristic time. Because the exponential growth  
produces so much radiation, we must assume that the abortion takes  
place after only about a femtosecond. In this picture, therefore,  
the observed pulse really consists of an exceedingly rapid  
exponential rise of duration a femtosecond or so, followed by a  
relatively slow fall lasting tens of picoseconds, for which the  
dissipative mechanism is responsible.

\bigskip\noindent
{\bf V. ~Conclusions}

It has been argued in the literature that vacuum radiation cannot be  
the source of sonoluminescence because the static Casimir energy is  
so small. Without entering into the controversy over how big the  
static Casimir energy is for a spherical bubble [18], we believe  
that, as illustrated by our model (which, after all, correctly  
reproduces the static Casimir energy in the case of parallel plates)  
there are two additional factors that ought to be taken into  
account: ~1) the Casimir effect arises essentially from the coupling  
of the electromagnetic field to a boundary. When that boundary is  
moving, the field is coupled to a time-dependent source, which in  
and of itself leads to the production of energy; and 2) if this  
time-dependent coupling gives rise to unstable modes, as it does in  
our model, then an unexpectedly large amount of energy can be  
produced.

The present work raises a number of issues for further  
investigation. Perhaps most important is tightening up the  
approximate treatment we have given for the classical solutions in  
the case of the collapsing bubble. It would be preferable to have a  
method of analysis that would conclusively demonstrate whether the  
period of exponential growth exists, and whether the model contains  
not only a mechanism for turning on the pulse but also for turning  
it off. Failing that, it will probably be necessary to include  
additional physics having to do with the kind of dissipation  
mechanism discussed above, capable of damping the vacuum radiation  
to a level compatible with what is seen experimentally. If this is  
the case, then the shape of the sonoluminescent pulse should exhibit  
a very rapid rise followed by a much slower decay.

Within the context of the expansion employed in this paper, as might  
be expected one finds that the higher orders in $\epsilon$ become  
progressively more complicated. We have examined the next  
non-trivial term, which is $\epsilon^5$, and we have verified that  
it does not qualitatively change the exponential behavior found in  
order $\epsilon^3$. We have not checked, however, that the  
$\epsilon^5$ contribution is numerically small compared to the  
$\epsilon^3$ contribution.

It would be useful as well to quantize the electromagnetic field in  
the presence of the collapsing bubble, much as we did for the  
simpler purely time-dependent case. We believe that any exponential  
behavior in the classical system will persist in its quantum  
counterpart, but having the explicit expression for quantum vacuum  
radiation would allow one to compare the details of the photon  
spectrum with experiment. It would also be useful, for numerical  
work, to have a good analytical approximation to $R(t)$.

Ultimately, electromagnetic radiation can be produced only by  
charges in motion. In the case of sonoluminescence, whether those  
charges effectively reside at the boundary of the bubble, as we  
contend in this paper, or within the gas inside the bubble, is a  
question that still awaits definitive resolution.

\bigskip\bigskip\noindent{\bf Acknowledgements}

We thank Matt Lippert and Andreas Nyffeler for help in the early  
stages of this work. We are grateful to Bob Apfel and Jeff  
Ketterling for much enlightenment about the experimental aspects of  
sonoluminescence. This research was supported in part by DOE grant   
\#DE-FG02-92ER-40704.

\bigskip\noindent
{\bf Appendix: ~More about $F\tilde{F}$}

In this Appendix we add a few remarks about the properties of  
$F\tilde{F}$. The $f(x) F\tilde{F}$ interaction analyzed in the text  
was chosen as the most convenient form for coupling the  
electromagnetic field to the bubble boundary. Whether it correctly  
captures the essential physics of this coupling is a matter that  
will require further investigation.

Another way of writing $F\tilde{F}$ is just $2\vec{E} \cdot  
\vec{B}$. It is, apart from the familiar $F_{\mu\nu}F^{\mu\nu} = {1  
\over 2}(\vec{B}^2 - \vec{E}^2)$, the only Lorentz invariant that  
can be constructed from $\vec{E}$ and $\vec{B}$ by algebraic means.  
A term in the Lagrangian of the form $\theta  
F_{\mu\nu}\tilde{F}^{\mu\nu}$, with $\theta$ constant, has no  
physical consequence in an Abelian gauge theory such as  
electrodynamics, because, as noted in eq. (3), it is a total  
divergence. In a non-Abelian theory (such as quantum chromodynamics  
or the electroweak theory) this term is still a divergence, but it  
nevertheless gives rise to non-perturbative physical effects because  
of the non-trivial topological structures, called instantons, that  
exist in such theories.

The anomalous divergence of the $U(1)$ axial vector current is  
proportional to $F\tilde{F}$. This term is directly responsible for  
the decay of the $\pi^0$ meson into $2$ gamma rays, which is its  
dominant decay mode.

In string theory, because of a property called $S$-duality [20,21],  
two seemingly different theories can in fact be equivalent. This can  
be very useful because often one of the two theories is strongly  
coupled (and therefore interactable) whereas the other is weakly  
coupled. As it turns out, the system we have been studying in  
connection with sonoluminescence exhibits a simple form of  
$S$-duality. To see this, it is useful to consider a slightly more  
general Lagrangian:

$$
{\cal L} = - {1 \over 4} [\varphi_1(x) F_{\mu\nu}F^{\mu\nu} +  
\varphi_2(x) F_{\mu\nu} \tilde{F}^{\mu\nu}] ~.\eqno(A1)
$$

\bigskip\noindent
In the text we had $\varphi_1 = 1$ and $\varphi_2 = f(x)$. In string  
theory, $\varphi_1(x)$ would be related to the dilaton, whereas  
$\varphi_2(x)$ is known as the axion field.

This system is governed by two sets of equations:

$$
\partial_{\mu}\tilde{F}^{\mu\nu} = 0 \eqno(A2)
$$

\noindent
and

$$
\partial_{\mu} [\varphi_1 F^{\mu\nu} + \varphi_2 \tilde{F}^{\mu\nu}]  
= 0 ~. \eqno(A3)
$$

\bigskip\noindent
If we express $F_{\mu\nu}$ in the usual way as

$$
F_{\mu\nu} = \partial_{\mu} A_{\nu} - \partial_{\nu} A_{\mu} ~,  
\eqno(A4)
$$

\noindent
then (A2) is an identity, while (A3) is a dynamical equation  
obtained by varying the Lagrangian with respect to $A_{\mu}$.

We define

$$
\tilde{G}_{\mu\nu} = \varphi_1 F_{\mu\nu} + \varphi_2  
\tilde{F}_{\mu\nu} ~.\eqno(A5)
$$

\bigskip\noindent
In Minkowski space, as is easily shown, the dual of a dual is the  
negative of the original tensor. Therefore

$$
G_{\mu\nu} = - \varphi_1 \tilde{F}_{\mu\nu} + \varphi_2 F_{\mu\nu}  
~.\eqno(A6)
$$

\bigskip\noindent
We can invert these relationships to obtain $F$ and $\tilde{F}$ in  
terms of $G$ and $\tilde{G}$, and then substitute them into ${\cal  
L}$:

$$
{\cal L} = - {1 \over 4} ({1 \over \varphi_1^2 + \varphi_2^2})  
\left\{ - \varphi_1 G_{\mu\nu}G^{\mu\nu} + \varphi_2  
G_{\mu\nu}\tilde{G}^{\mu\nu} \right\} ~.\eqno(A7)
$$

\bigskip\noindent
If we set

$$
G_{\mu\nu} = \partial_{\mu} B_{\nu} - \partial_{\nu} B_{\mu}  
\eqno(A8)
$$

\noindent
and vary ${\cal L}$ with respect to $B_{\mu}$, we obtain

$$
\partial_{\mu} [{- \varphi_1 \over \varphi_1^2 + \varphi_2^2}  
G^{\mu\nu} + {\varphi_2 \over \varphi_1^2 + \varphi_2^2}  
\tilde{G}^{\mu\nu}] = 0 \eqno(A9)
$$

\noindent
and also

$$
\partial_{\mu} \tilde{G}^{\mu\nu} = 0 \eqno(A10)
$$

\bigskip\noindent
as an identity. It is straightforward algebra to show that (A9) is  
the same as (A2), and (A10) is, by definition, the same as (A3).  
Thus the physical content of the Lagrangian (A7) is the same as  
(A1), but the dynamical equation in one case is an identity in the  
other case, and vice versa. We see that if we can solve a system  
with sources $(\varphi_1, \varphi_2)$, then by duality we  
automatically obtain a solution with sources $({- \varphi_1 \over  
\varphi_1^2 + \varphi_2^2}, {\varphi_2 \over \varphi_1^2 +  
\varphi_2^2})$. It is not clear, however, whether practical use can  
be made of this observation in the case of sonoluminescence.

\bigskip\noindent
{\bf References}

\bigskip\noindent
[1] D.F. Gaitan, L.A. Crum, C.C. Church and R.A. Roy, J. Acoust.  
Soc. Am. {\bf 91}, 3166 (1992).

\noindent
[2] B.P. Barber, et al., Phys. Rev. Lett. {\bf 72}, 1380 (1994);  
R.G. Holt and D.F. Gaitan, Phys. Rev. Lett. {\bf 77}, 3791 (1996).

\noindent
[3] B. Gompf, et al., Phys. Rev. Lett. {\bf 79}, 1405 (1997); R.A.  
Hiller, S.J. Putterman, and K.R. Weininger, Phys. Rev. Lett. {\bf  
80}, 1090 (1998).

\noindent
[4] B.P. Barber and S.J. Putterman, Nature {\bf 352}, 318 (1991).

\noindent
[5] R.A. Hiller, S.J. Putterman, and B.P. Barber, Phys. Rev. Lett.  
{\bf 69}, 1182 (1992).

\noindent
[6] B.P. Barber, et al., Phys. Rev. Lett. {\bf 72}, 1380 (1994).

\noindent
[7] R.A. Hiller, et al., Science {\bf 266}, 248 (1994).

\noindent
[8] D. Lohse, et al., Phys. Rev. Lett. {\bf 78}, 1359 (1997); T.J.  
Matula and L.A. Crum, Phys. Rev. Lett. {\bf 80}, 865 (1998).

\noindent
[9] See, for example, J.D.N. Cheeke, Can. J. Phys. {\bf 75}, 77  
(1997); B.P. Barber, et al., Phys. Rep. {\bf 281}, 65 (1997).

\noindent
[10] B.P. Barber and S.J. Putterman, Phys. Rev. Lett. {\bf 69}, 3839  
(1992); C.C. Wu and P.H. Roberts, Phys. Rev. Lett. {\bf 70}, 3424  
(1993), and Proc. Roy. Soc. A {\bf 445}, 323 (1994); L. Frommhold  
and A.A. Aitchley, Phys. Rev. Lett. {\bf 73}, 2883 (1994); A. K.  
Evans, Phys. Rev. E{\bf 54}, 5004 (1996); W.C. Moss, et al., Science  
{\bf 276}, 1398 (1997); K. Yasui, Phys. Rev. E{\bf 56}, 6750 (1997);  
L.P. Csernai and Z. Lazar, Phys. Lett. A{\bf 235}, 291 (1997); P.  
Mohanty and S.V. Khare, Phys. Rev. Lett. {\bf 80}, 189 (1998).

\noindent
[11] J. Schwinger, PNAS {\bf 89}, 4091 (1992); ibid., {\bf 89},  
11118 (1992); ibid., {\bf 90}, 958, 2105, 4505 and 7285 (1993);  
ibid., {\bf 91}, 6473 (1994).

\noindent
[12] C. Eberlein, Phys. Rev. A{\bf 53}, 2772 (1996); Phys. Rev.  
Lett. {\bf 76}, 3842 (1996).

\noindent
[13] E. Sassaroli, Y. Srivastava, and A. Widom, Phys. Rev. A {\bf  
50}, 1027 (1994); Nucl. Phys. B, {\bf Suppl. 33C}, 209 (1993); E.  
Sassaroli, et al., hep-ph/9805479.

\noindent
[14] A. Chodos, in "{\it Neutrino Mass, Dark Matter, Gravitational  
Waves, Monopole Condensation and Light Cone Quantization}," B.N.  
Kursunoglu, S.L. Mintz and A. Perlmutter, eds., Plenum Plress, New  
York and London, 1996, p. 371.

\noindent
[15] See also G. Calucci, J. Phys. A {\bf 25}, 3873 (1992).

\noindent
[16] This equation has appeared in a variety of other contexts. See,  
e.g., B. Nodland and J.P. Ralston, Phys. Rev. Lett. {\bf 78}, 3043  
(1997).

\noindent
[17] H.B.G. Casimir, Proc. K. Ned. Akad. Wet. {\bf 51}, 783 (1948).

\noindent
[18] K. A. Milton and Y.J. Ng, Phys. Rev. E {\bf 55}, 4207 (1997);  
C.E. Carlson, C. Molina-Paris, J. Perez-Mercader and M. Visser,  
Phys. Rev. D {\bf 56}, 1262 (1997) and Phys. Lett. {\bf B395}, 76  
(1997); C. Molina-Paris and M. Visser, Phys. Rev. D {\bf 56}, 6629  
(1997); K. A. Milton and Y.J. Ng, Phys. Rev. {\bf E57}, 5504 (1998);  
V.V. Nesterenko and I.G. Pirozhenko, JETP Lett. {\bf 67}, 445  
(1998).

\noindent
[19] S. Liberati, et al., quant-ph/9805023 and quant-ph/9805031.

\noindent
[20] A. Sen, Int. J. Mod. Phys. {\bf A9}, 3707 (1994).

\noindent
[21] J. Schwarz, lectures at TASI (1996).

\end